\documentclass[aps,prl,showpacs,twocolumn,superscriptaddress,floats]{revtex4} 

\usepackage{graphicx}
\usepackage{bm}
\usepackage{epsfig}
\input epsf

\usepackage{color,hyperref}
\definecolor{blue}{rgb}{0,0.0,0.9}
\definecolor{green}{rgb}{0.0,0.9,0.0}

\begin{document}

\newcommand{\nsl}[1]{\rlap{\hbox{$\mskip 1 mu /$}}#1}
\preprint{APS/123-QED}

\title{The effect of a velocity barrier on the ballistic transport of Dirac fermions}

\author{A. Concha}
\author{Z. {Te{\v s}anovi{\'c}}}
 
\affiliation{Department of Physics and Astronomy, Johns Hopkins University, Baltimore, Maryland 21218}


\begin{abstract}
We propose a novel way to manipulate the transport properties of massless 
Dirac fermions by using velocity 
barriers, defining the region in which the Fermi velocity, $v_{F}$, 
has a value that differs from the one in the surrounding background.
The idea is based on the fact that when waves travel accross different media, there are boundary 
conditions that must be satisfied, giving rise to Snell's-like laws. 
We find that the transmission through a velocity barrier is highly anisotropic, and that perfect 
transmission always occurs at normal incidence. When $v_{F}$ in the barrier is larger that the 
velocity outside the barrier, we find that a critical transmission angle exists, 
a Brewster-like angle for massless Dirac electrons. 
\end{abstract}


\maketitle
When reading this paper, one is using the fact 
that the speed of light in vacuum is different from 
the speed of light in various parts of your eyes \cite{landau1984}. 
That difference allows our eyes and 
other optical devices to focus light in a very simple but efficient way.
In general, when a physical object crosses a boundary, it must 
follow certain rules, regardless of its 
particle or wave-like behavior.
Those rules are typically various incarnations of the well-known Snell's law for optics.

In optics, the Snell's law is the natural outcome of Fermat's principle: light follows the 
path of least time \cite{landau1984,arnold1989}. 
Similar  laws are found in all known oscillatory phenomena \cite{fetter1980}. This 
relation can also 
be found in classical mechanics in the standard problem of scattering by a constant potential 
barrier \cite{landau1989}. In this 
case, this law appears as the consequence of conservation of linear momentum in the direction parallel to the barrier 
and overall energy conservation.
It also appears in quantum systems, and the prediction of relations analogous to Snell's law would be of utmost 
importance because, as happens in optics, it will allow us to control the focusing of electrons, opening paths 
for new nanodevices \cite{cheianov2007}.

With the successful preparation of graphene -- a single layer of 
graphite \cite{novoselov2005,kim2005} -- a new route to 
test long standing predictions made in quantum electrodynamics became possible \cite{greiner2008}. This new material has 
also opened new ways to fabricate nanodevices that take advantage of the multiple 
exotic characteristics and novel phenomena shown by graphene, such as unimpeded penetration 
of quasi-particles through p-n junctions \cite{katsnelson2006,shytov2008,young2009}, the 
possible control of pseudospin number
(valleytronics) \cite{rycerz2007}, or metrology applications such as the 
measurement of the fine structure constant \cite{nair2008}.

 Quite recently, it has been argued that electron super-collimation 
could be achieved in graphene by using a potential super-lattice \cite{park2008b}. This approach requires 
a careful control of the potential 
barriers. Given that the system will essentially be one dimensional, it would be difficult 
to avoid the effects of disorder, although it is well-known that massless Dirac fermions 
are not quite as susceptible to potential barriers as their Schr\"{o}edinger cousins.

In this paper, we describe a novel, velocity barrier approach to collimation and manipulation of
beams of massless Dirac particles. This approach
is based on the fact that the above super-lattice will produce an anisotropic 
velocity renormalization, making the effective velocity on 
the vertical coordinate  $\left(v_{y}\right)$  smaller than 
the original Fermi velocity $\left(v_{F}\right)$ for 
excitations in clean graphene. Thus, due to momentum conservation, it 
can be argued that only electrons 
that are close to normal incidence will survive the scattering 
with the super-lattice. All electrons with 
momentum  far from normal incidence will be deflected.
In this  argument, the role of momentum conservation in the direction 
perpendicular to the barrier is fundamental, as is the case for photons. 
Momentum in the direction 
parallel to the barrier is not conserved. Thus, if we force it 
to change, the outcome obtained by the above 
argument will remain. This 
is precisely the case when a velocity barrier is used.
\begin{figure}[t]
\leavevmode
\epsfxsize=8.6cm
\epsfbox{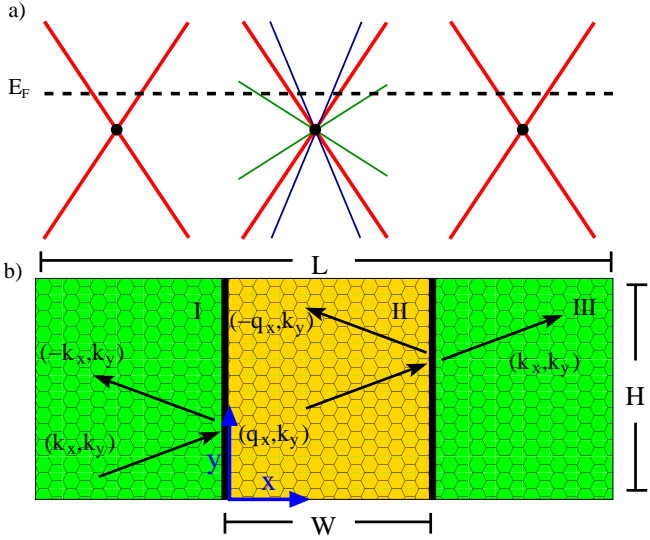}
\caption{\label{FIG1}(Color online). a) Schematic diagram of the low-energy spectrum of Dirac quasi-particles 
 when a velocity barrier is present. The three diagrams in a) show how the Fermi velocity 
$v_{F}$ (the slope of the Dirac cone) changes as a function of $x$. The green and blue Dirac cones correspond
to $v_{F}<v_{F}^{0}$ and $v_{F}>v_{F}^{0}$ respectively.
b) Setup needed to test the predicted effects. We set the Fermi velocity $v_{F}^{0}$ of 
the system for $x<0$ and $x>W$ to one. $v_{F} $ is the Fermi velocity 
for $0<x<W$, rendering the system non-homogeneous.  The wave vectors used to 
find the transmission matrix solution of this problem are shown in black.}
\noindent
\end{figure} 

On the experimental front, a velocity barrier can be implemented 
in several ways. For example, one could stretch 
a small region of a graphene sheet \cite{castroneto2009}, use 
super-lattices \cite{park2008c,gibertini2009} or vary 
the interactions with the medium around the graphene layer \cite{jang2008,bostwick2007}. 
Here, we solve a generic 
problem in which the Fermi velocity has been modified to 
form what we call a velocity barrier. We must emphasize 
that our results are completely independent of the method used 
to modify the Fermi velocity, provided there is 
no gap opening in the system.

The proposed setup is shown in Fig.\ref{FIG1}, in which Dirac fermions move with a group velocity given by:
$$
v_{eff}(x) = \left\{ \begin{array}{rl}
      v_{F}^{0}         , &\mbox{Region $\bf{I}$, $x<0$}   \\ 
      v_{F}     , &\mbox{Region $\bf{II}$, $0<x<W$ } \\
      v_{F}^{0}         , &\mbox{Region $\bf{III}$, $x>W$}
       \end{array} \right. $$
We will set $v_{F}^{0}$ to unity, thus the only relevant quantity will be $v_{F}$, the Fermi velocity 
inside the barrier, expressed in units of the Fermi velocity far from the barrier. $W$ is the barrier width.

In the absence of external potentials, quasi-particle excitations in graphene obey the 
Dirac equation \cite{wallace1947}:
\begin{eqnarray}
v_{eff}\vec{\sigma}\cdot\vec{p}\psi=E\psi
\label{deq}
\end{eqnarray}
where $\vec{\sigma}=(\sigma_{x},\sigma_{y})$ is in standard Pauli matrix notation. For convenience, we 
define $\hbar=1$. This equation has a generic chiral 
solution around the Dirac point $\vec{K}$, which can be written (after a gauge transformation) in 
momentum space as: 
$$\psi(\vec{k})=\frac{1}{\sqrt{2}} \pmatrix{   1   \cr
                                               s e^{i\theta_{\bm{k}}}    \cr
                                                     }$$
where $\theta_{\bm{k}}=\arctan(k_{x}/k_{y})$ is the angle defined in momentum space. $s=\pm$ indicates 
the chirality of the solution which, for the case of graphene like structures, is 
associated with the current $\left(\vec{J}=e v_{F}\psi^{\dagger} \vec{\sigma} \psi \right)$ and not with the handedness of the 
system. Note that in the problem discussed in this paper, chirality will not play an important 
role and we are free to set  $s=1$. We have assumed that 
the barrier is smooth compared with the lattice spacing of the underlying physical system, such 
that no $\bf{K}$ $\bf{K}'$ valley mixing will occur.

We can write the general solution for this scattering problem in terms of the 
incident and reflected waves. In region $\bf{I}$ we have that:
$$\psi_{\bf{I}}(\vec{r})=\frac{1}{\sqrt{2}} \pmatrix{   1   \cr
                                                    e^{i\phi}    \cr} e^{i(k_{x}x+k_{y}y)}
                                                     +
                     \frac{r}{\sqrt{2}} \pmatrix{   1   \cr
                                                    e^{i(\pi-\phi)}    \cr} e^{i(-k_{x}x+k_{y}y)}
$$, 
\begin{figure}[t]
\leavevmode
\epsfxsize=7.5cm
\epsfbox{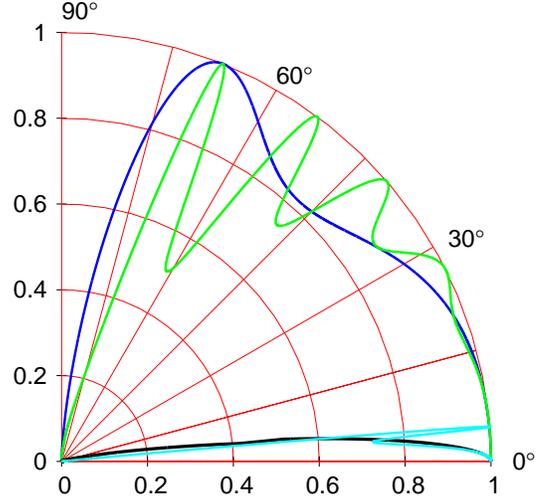}
\caption{\label{FIG2}(Color online). Transmission probability for different values of the control parameters $E_{F}$ and $v_{F}$.
Green and blue lines correspond to the case in which the effective Fermi 
velocity inside the barrier $0<x<W$ is $1/2$. Blue corresponds to $E_{F}$ 
equal to $20$ meV and green to $100$ meV. 
Black and cyan lines correspond to the cases in which the effective Fermi 
velocity inside the barrier $0<x<W$ is $10$. Black corresponds to $E_{F}$ 
equal to $20$ meV and cyan to $100$ meV.  The barrier width is $W=350$ nm for all 
curves. Notice that, for large $v_{F}$, there is a critical angle (Brewster angle) for which no 
transmission is possible .  
}
\noindent
\end{figure} 
where $\phi=\arctan(k_{y}/k_{x})$, $k_{x}=k_{F}\cos(\phi)$, and $k_{y}=k_{F}\sin(\phi)$. 
In region $\bf{II}$ the solution can be constructed in a similar fashion as:

$$\psi_{\bf{II}}(\vec{r})=\frac{a}{\sqrt{2}} \pmatrix{   1   \cr
                                                    e^{i\theta}    \cr} e^{i(q_{x}x+k_{y}y)}
                                                     +
                     \frac{b}{\sqrt{2}} \pmatrix{   1   \cr
                                                    e^{i(\pi-\theta)}    \cr} e^{i(-q_{x}x+k_{y}y)}
$$, 
where $\theta=\arctan(k_{y}/q_{x})$, and $q_{x}=\sqrt{\left(\frac{E}{v_{F}}\right)^{2}-k_{y}^{2}}$.

\begin{figure}[t]
\leavevmode
\epsfxsize=7.5cm
\epsfbox{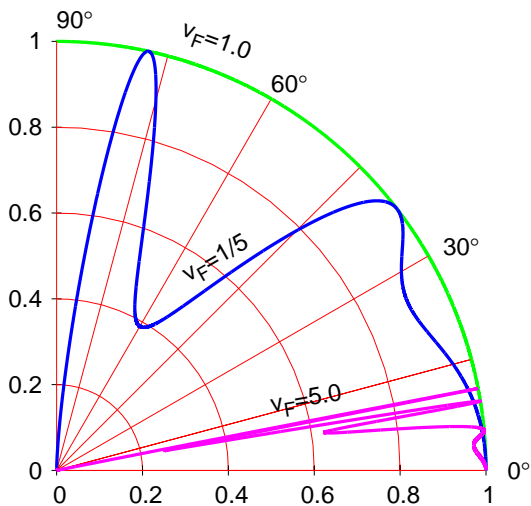}
\caption{\label{FIG3}
(Color online). Transmission probability at $E_{F}=100$ meV, $W=350$ nm and velocities inside the barrier
$v_{F}=1$, $1/5$, and $5$ (green, blue, and pink lines respectively).  
Resonances are apparent at $W q_{x}=n \pi$, with $n=0,1,2,...$. Note that  for $v_{F}>1$ no 
transmission is allowed if $\phi>\phi_{cr}$. 
}
\noindent
\end{figure}

\begin{figure}[th]
\leavevmode
\epsfxsize=8.6cm
\epsfbox{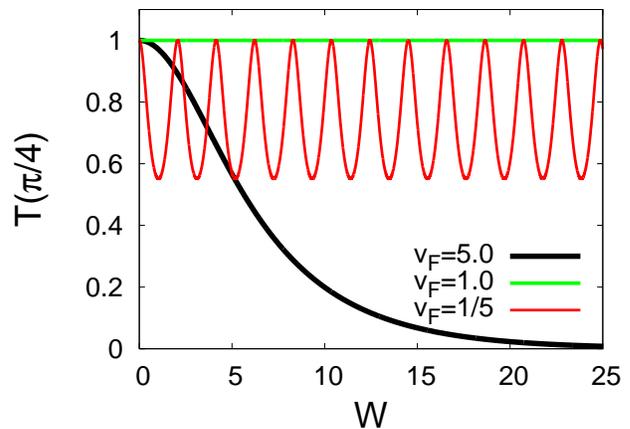}
\caption{\label{FIG4}
(Color online). Transmission probability as a function of the width $W$ for fixed energy $E_{F}=100$ meV and 
angle $\phi=\pi/4$. Curves in green, red, and black correspond to  velocities $v_{F}=1$, $1/5$, and $5$ , respectively, inside 
the barrier.
}
\noindent
\end{figure}

For the transmitted wave we have:
$$\psi_{\bf{III}}(\vec{r})=\frac{t}{\sqrt{2}} \pmatrix{   1   \cr
                                                    e^{i\phi}    \cr} e^{i(k_{x}x+k_{y}y)}                    
$$
Thus, in principle we have to solve the scattering problem using the transmission matrix approach. In this problem,
 the correct boundary conditions to be imposed at $x=0$ and $x=W$ are:
\begin{eqnarray}
\psi_{\bf{I}}(0^{-})&=&\sqrt{v_{F}}\psi_{\bf{II}}(0^{+})\\
\psi_{\bf{II}}(W^{-})&=&\frac{1}{\sqrt{v_{F}}}\psi_{\bf{III}}(W^{+}).
\label{bc}
\end{eqnarray}
These boundary conditions are a consequence of the conservation of local current at the
interfaces.
Solving for the coefficients $a,b,r$, and $t$, we find for the reflection coefficient:
\begin{eqnarray}
r=\frac{e^{i \phi } \sin(W q_{x})[\sin(\phi)-\sin(\theta)]}{
\cos(W q_{x})\cos(\theta )\cos (\phi )+i \sin(W q_{x})[\sin(\theta)\sin(\phi) -1]}
\label{req}
\end{eqnarray}
Figs.(\ref{FIG2}) and (\ref{FIG3}) show the angular dependence of the transmission probability $T=1-|r|^2$. It 
is important to note that $T\left(\phi\right)=T\left(-\phi\right)$. Taking advantage of that symmetry, we plot our results in the 
interval $\phi \in [0,90^{0}]$ degrees. 
Furthermore, for the case of massless particles at normal incidence we find 
that $T\left(0\right)=1$, indicating perfect transmission at normal incidence regardless of the value of $v_{F}$. 
 The existence of peaks that reach perfect transmission 
at specific angles is characteristic of resonant behavior in this system. This can be readily 
checked by analyzing the zeros of $r$ that 
correspond to $\sin(Wq_{x})=0$, producing resonances at $Wq_{x}=n\pi$ for integer values of $n$ (in 
other words, when the barrier becomes transparent). A second peculiarity appears when $v_{F}>1$. As shown 
by the black and cyan lines in Fig.(\ref{FIG2}) and in pink in Fig.(\ref{FIG3}), transmission is 
prohibited for any angle larger than $\phi_{c}$. In Figs.(\ref{FIG3}) and (\ref{FIG4}), where we show the transmission at a 
fixed energy for different velocities and width ($W$), the existence of a critical angle is apparent when $v_{F}>1$.   
It is also clear from Fig.(\ref{FIG2}) that the only effect of increasing the energy is to increase the number of
resonant peaks in $T(\phi)$. In fact, as the energy of the incident Dirac fermions is increased, more ballistic 
channels will be opened.
The critical angle $\phi_{c}$ is analogous to the called Brewster angle in optics \cite{landau1984}. From 
the definition of the refracted angle:
\begin{eqnarray}
\theta=\pm\arctan\left(\frac{sign\left[E_{F}\right]\sin\phi}{\sqrt{\left(\frac{1}{v_{F}}\right)^2-\sin^{2}\phi}}\right)
\label{refracted}
\end{eqnarray}
we can see that the critical angle will be given by $\sin^{2}\phi_{c}=v_{F}^{-2}$, which has solutions 
only if $v_{F}>1$, independent of $E_{F}$.
  
We can make a link to the standard Klein's paradox by pointing out that Eq.(\ref{req}) has 
the same functional form of the reflection of Dirac fermions against a potential barrier with a 
rectangular shape \cite{katsnelson2006}. 
The only difference is that the change in the Fermi velocity is encoded in the definition of $q_{x}$. This 
simple observation allows us to compare the effect of a velocity barrier in terms of a square 
potential. The dynamical variable $k_{y}$ is invariant due to the translational symmetry in the vertical 
direction. Therefore, it is enough to fix the second dynamical variable $q_{x}$ equal in both cases, and 
then solve for the potential in order to produce the same transmission coefficients in both problems.
It is straightforward to show that such a potential will be: 
\begin{eqnarray}
V\left(E\right)=E \pm \frac{|E|}{v_{F}}
\label{pot}
\end{eqnarray}
By using this analogy, we have obtained an energy dependent potential, which renders new phenomenology of our velocity 
barrier. Such energy dependent potentials have been used in nuclear physics \cite{arellano1989}, and in a 
different context in attempts to generalize the uncertainty principle for 
high energy physics \cite{saavedra1981}.  
\begin{figure}[th]
\leavevmode
\epsfxsize=8.6cm
\center
\epsfbox{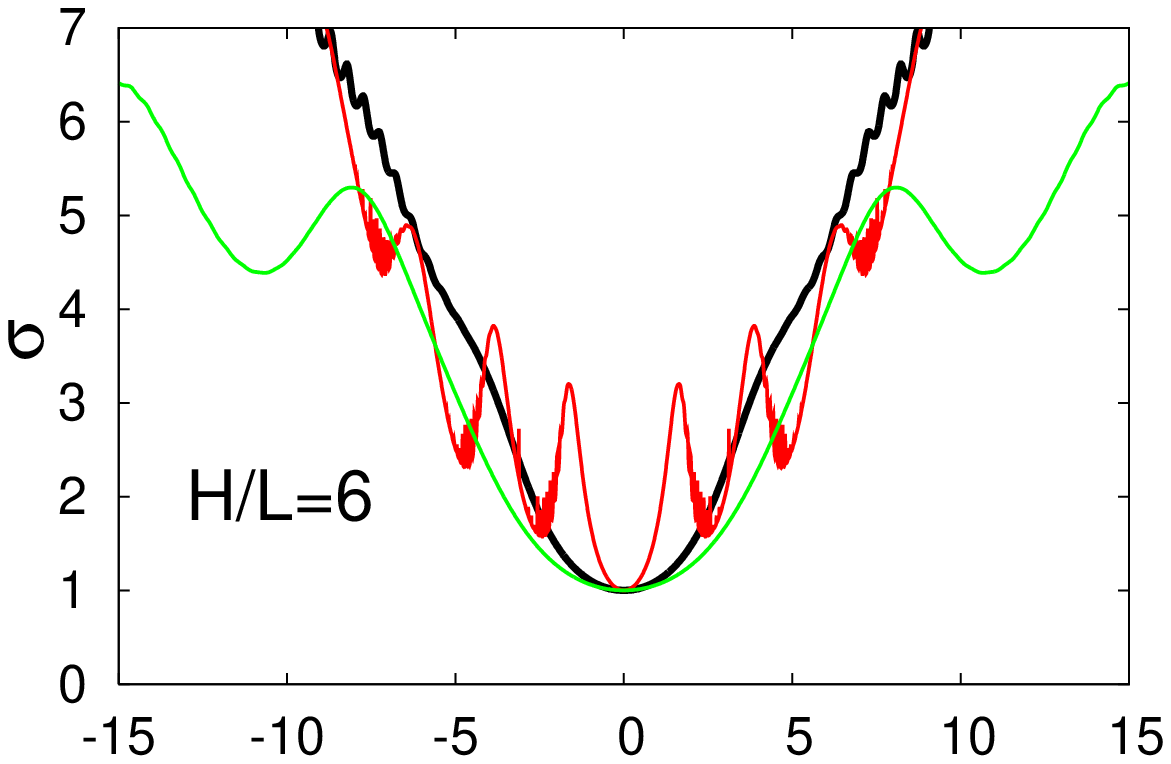}
\epsfxsize=8.6cm
\epsfbox{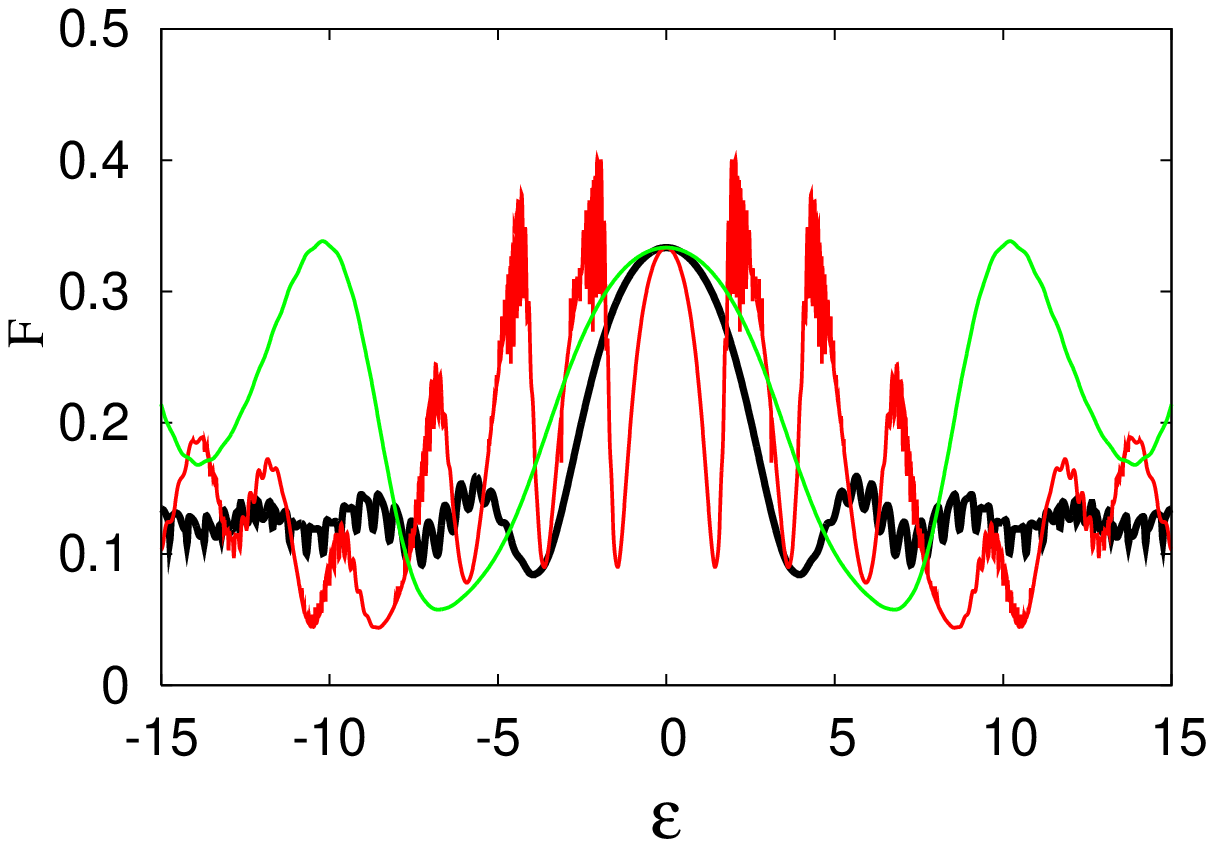}
\caption{\label{FIG5}
Conductivity  $\sigma$ in units of $2e^{2}/h$ and Fano factor $F$ as a function of the energy of the incoming Dirac 
fermions: Green, Red, and black correspond to $v_{F}$ $5$, $0.2$, and $1.0$ respectively. $\epsilon$ is $e V H/h v_{F}$ where $V$ is the gate voltage.
}
\noindent
\end{figure} 
We have also computed the effects of a velocity barrier on two directly measurable quantities for this system: 
conductivity and Fano factor \cite{2008DiCarlo,2008Danneau}. In the ballistic approximation both quantities are computed 
using the following formulae:
\begin{eqnarray}
\sigma=\left(\frac{L}{H}\right)\sum_{n=-\infty}^{\infty} T_{n} ,\;\; F=\frac{\sum\limits_{n=-\infty}^{\infty}T_{n}(1-T_{n})}{\sum\limits_{n=-\infty}^{\infty}T_{n}},\label{fano}
\end{eqnarray}
The transmission of each channel $T_{n}$ depends on a  phase factor, $\alpha$, which has different values 
for different boundary conditions. In Fig.(\ref{FIG5}), we have used $\alpha=1/2$, corresponding 
to an infinite mass lateral confinement \cite{1987Berry,2006Tworzy}. We have also checked that, in the wide ribbon limit,
 our results stay the same for other boundary conditions.

Our computation shows that $\sigma$ and $F$ are indeed sensitive to the presence of  a velocity barrier. As 
expected from Eq.(\ref{pot}), the zero energy values of conductance and Fano factor remain the 
same. However, their values will drastically change for a small bias voltage. In Fig.(\ref{FIG5}) the 
red line indicates the values for $\sigma$ and $F$ for $v_{F}=1/5$, showing that conductance 
increases by a factor $3$ at $\epsilon\sim 1.5$, as well as large amplitude oscillations close to the Dirac point. For the 
case in which $v_{F}=5$ (green) few momenta can penetrate the barrier due to the existence of a Brewster angle in this case. 
In turn, the conductance is greatly reduced for energies away from the Dirac point. In this case, the Fano factor increases 
and reaches values that indicate departure from the classical diffusive limit $F=1/3$. In particular, the broad 
oscillation in $F$ in this case should be experimentally 
measurable \cite{2008DiCarlo,2008Danneau}.      
 
In this paper we have proposed a novel route to control
the electric transport of massless Dirac fermions, based on 
experimental control \cite{park2008c,gibertini2009} of velocity barriers. 
Our main results demonstrate that it is possible, in principle, to manipulate the 
transmission properties of a system described by a Dirac equation 
by controlling the Fermi velocity. The similarity of the Fermi velocity 
to the role of the refractive index 
in optics naturally results in an effective Brewster 
angle, which bodes well for ultimate construction of waveguides and related devices.
We have also shown that the Fano factor and the conductivity of this system can be modified 
by a velocity barrier, producing strong oscillations as we move away from the Dirac point.     
It would be interesting to produce exact simulations in order 
to study the precise energy window in which 
these effects should be observable. 

This work was supported in part by the 
NSF under Grant No. DMR-0531159.

\end{document}